\documentclass[preprint,eqsecnum,aps]{revtex4}

\usepackage{graphicx}%
\usepackage{dcolumn}

\begin{document}


\title[]{ 
PROBING PORES USING ELEMENTARY QUANTUM MECHANICS}
\author{{\sc Seungoh Ryu}}
\affiliation{%
Schlumberger-Doll Research, 
Old Quarry Road, Ridgefield, CT 06877}%
\begin{abstract}
{\bf The relaxation of polarized spins in a porous medium has been utilized as a probe of its structure. 
We note that the governing diffusion problem has a close parallel to that of a particle in a box,
an elementary Quantum mechanics toy model. 
Following the spirits of ``free electron" model, we use generic properties of the eigen spectrum to
understand features common to a wide variety of pore geometry, consistent with large scale numerical
simulations and experimental data. 

}
\end{abstract}

\maketitle

\centerline{{\bf INTRODUCTION}}
\vskip 0.08in
\noindent
Relaxation of polarized spins by paramagnetic impurities embedded in pore matrix provides a valuable probe into
its structure, especially when elaborate techniques such as MRI imaging are not easily applicable. Exploration
for hydrocarbon inside a bore hole is one such case where it is critical to extract limited, but pertinent
information in a timely manner from measurements often done with a hand tied.  Over the past years, a lot of
progress has been made both in development of magnetic resonance tools$^1$ as well as theoretical
aspects$^2$ geared specifically toward extracting surface-to-volume ratio from the observed
magnetization evolution.  There are several more sophisticated MR techniques now being routinely
applied to rocks in the laboratories and thus potentially much more detailed information on the structure of the
pore space is available.$^3$ In addition to the pore structure which controls practically important
properties such as ite permeability, one is often interested in the details of pore filling fluids or gas such
as how is is made up out of water and oil, diffusion constants of its constituents, and wetting condition. As
our scope begins to embrace this degree of complexity, it becomes essential to bring some concepts  up to date
since sorting out  the complexity of the pore matrix from that of its content is a minimum requirement in such
endeavor. 

Although we are specifically motivated by MR applied to porous media, the underlying diffusion phenemenon under
complex boundary conditions occurs in a variety of {\em real life} situations.  To illustrate a useful point,
let us consider an extreme case: zombies wandering on top of a shopping mall whose fence has occasional openings
at the edge.  Depending on the size of the space, the speed of a zombie, and the frequency of the openings, the
depletion of its population occurs either in a homogeneous way or with more interesting profile in space.  The
conventional approach {\em sans} imaging relied on deducing these parameters from observing how the overall
population decays.

\vskip 0.08in
\centerline{{\bf QUANTUM MECHANICS}}
\vskip 0.08in
The analysis is simple if we take the continuum approach, treating the
population as a scalar variable, subject to the classic diffusion equation.$^{4,5}$  The diffusive current
$j_D$ arises from inhomogeneous density distribution $m$ via 
$j_D = - D \nabla m$. Conservation of local flux $\nabla \cdot j_D = - {d \over dt} m$ then leads to 
$ D \nabla^2 m = {d \over dt} m$ with the boundary condition $\hat {n} \cdot j_d = \rho m$ on the wall. 
Treating it as a boundary value problem, the problem is solved in terms of eigenmode expansion, as was studied
in detail by Brownstein and Taar for simple geometries years earlier. In this work, we employ an analogy to
Quantum mechanics by observing that 
$D \rightarrow {\hbar \over 2 m}$ and $t \rightarrow i t$ turns the equation for $m$ into the Schr\"odinger
eqaution for a particle confined to a {\em leaky} container with 
${\cal H} = - {\hbar^2 \over 2m }\nabla^2$. 
$m$ is identified with the wave function $\Psi$ of the QM particle in an imaginary time, but we note that the
$m$ is directly observed in our classical system, rather than $ |\Psi |^2$.  The Hamiltonian operator is simply
that of kinetic energy of the particle and the boundary condition is of mixed type, $[\hat {n} \cdot D \nabla +
\rho ] m = 0$.  The existence, orthonormality of eigenmodes are guarranteed with $\Psi_n$ satisfying 
$
{\cal H} \Psi_n = E_n \Psi_n.
$
By multiplying both sides with $\Psi_n^*$ and integrating over the pore volume, with the help of
the boundary condition, 
one can show 
\begin{equation}
\label{eq:en2}
\Big ( {\hbar^2 \over 2m }
  \int d^3r |\nabla \Psi_n |^2  +\rho \hbar
 \oint | d{\bf s} | | \Psi_n |^2 \Big) = E_n,
\end{equation}
which shows that $E_n \ge 0$ with equality allowed only for $\rho = 0$ with a uniform mode $\Psi_0 = {1 \over \sqrt{V}}$. 
The energy (relaxation rate) is broken down into two components: the kinetic energy part and the potential energy part which is localized in the surface region. 
For each eigenmode, we have a length scale 
$\ell_n = {\oint | d{\bf s} | | \Psi_n |^2  \over  \int d^3r |\nabla \Psi_n |^2}
$ with which we have
\begin{equation}
E_n =  {\hbar^2 \over 2m }
\int  d^3r |\nabla \Psi |^2  ( 1 +  {2m \rho \over \hbar} \ell_n ).
\end{equation}
Since $\ell_n \propto {{\cal S}\over V}{1\over k_n^2}$ decreases with increasing $n$, 
this clearly shows how the high (excited) modes become progressively independent of $\rho.$

From now on, we introduce the bra-ket notation, $\Psi_n({\bf r}) = <{\bf r} | n >$ for compactness. 
The eigenmodes form a complete set with 
$| n > \sum_{n=0}^\infty < n| = 1$ 
and the time evolution of an arbitrary state $| \Psi_i >$ is given by 
\begin{equation}
| \Psi_t> = \sum_n |n> e^{-{E_n\over \hbar} t} <n | \Psi_i >.
\end{equation}

\begin{figure}[p]
\includegraphics[width =3 in]{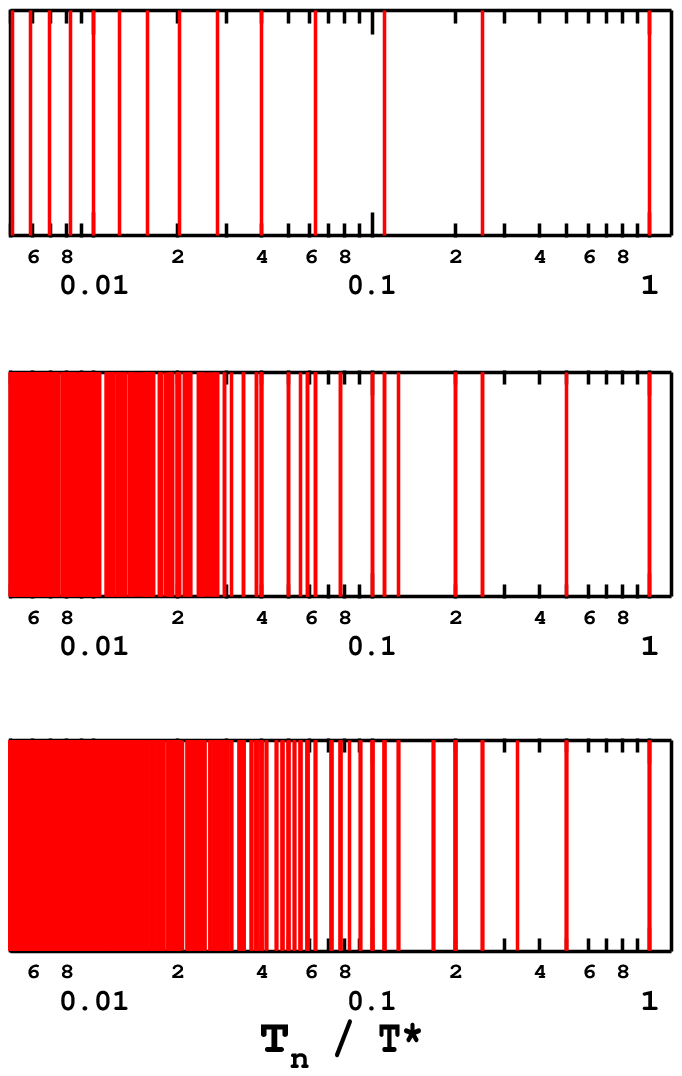} \includegraphics[width = 2. in]{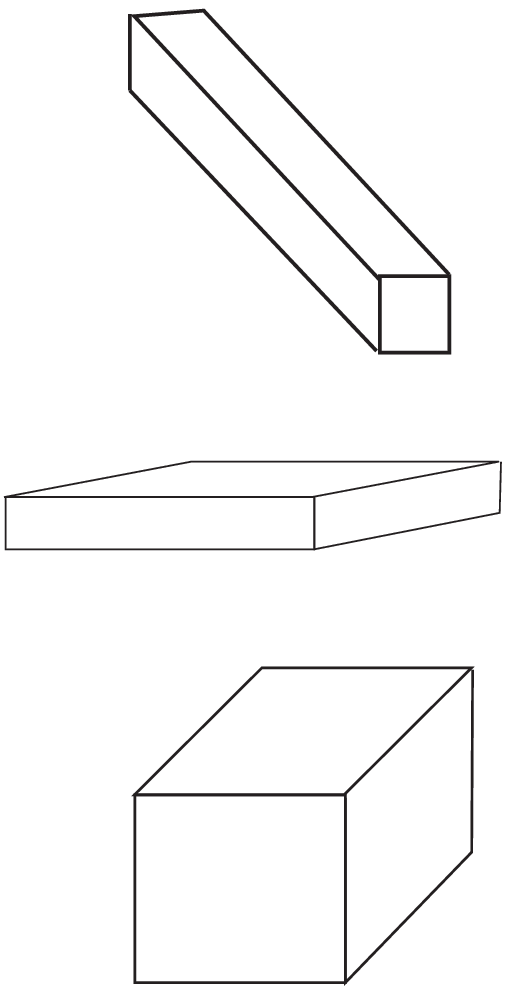}\\
\caption{Spectral ``fingerprints" of the boundary shape: For a given volume $L^3$ (porosity), we compare
three  distinct aspect ratio $({L_x\over L},{L_y\over L},{L_z\over L}) = (11.11,0.3,0.3)$(top:rod),
$(1.826,1.826,0.3)$ (middle:slab)  and $(1,1,1)$ (bottom:cube). $T^* \equiv L_x^2/D\pi^2$:nominal
diffusion time along the longest length. $\rho = 0$ is used.
 }
\label{fig:box}
\end{figure}

This is all very formal. The hard part lies in the specific  spectrum for a given boundary shape and
the strength of the sink. The distribution of eigenvalues encodes much information of the boundary
shape, i.e.~pore structure.$^6$ It is possible that there exists significant degeneracy, since different
shapes may lead to an identical spectrum.  Despite this, we anticipate that we can relate some
characteristics of the eigenspectrum to a given shape of the pore wall. Borrowing the concepts from
random matrix theory and Quantum chaos, we may study the correlation among the levels via  evaluating
${\cal P} (\epsilon_{nm})$ where $\epsilon = E_n - E_m.$ This function generally displays a universal
characterisitc for large $E_n$'s, while at small values (correlation among slow diffusion modes)
unique fingerprint of the boundary structure manifests itself. 
Note that $\rho$ acts as a knob taking the system from a Neunmann ($\rho \rightarrow 0$) to a Dirchlet
($\rho \rightarrow \infty$) boundary condition. This variation in  sink strength largely affect the
lowest eigen mode in a non-trivial manner while for higher levels, amounts only to a uniform phase shift by
$\sim {\pi \over 2}$.
The effect of a large scale pore shape on the spectrum can be most easily appreciated in the model
system of particle in a rectangle. There the spectrum follows ( in the $\rho\rightarrow 0$ limit for
simplicity):
\begin{equation}
{E_{\bf n} \over \hbar}  = {1 \over T_{\bf n}} = {\hbar \pi^2 \over 2 m}  [ ({n_x \over L_x})^2 +
({n_y
\over L_y})^2 + ({n_z
\over L_z})^2 ].
\end{equation}
One can easilty obtain the density of states from the surface area of an ellisoid in the ${\bf k}$ space as in
the free electron model. Some representative spectra are shown in Figure \ref{fig:box} in which we varied the
aspect ratio of a rectangle while keeping its volume fixed at $L^3 = 1$ and $\rho = 0.$
Qualitatively, the shape of the bouding boxes represents a 1D-, 2D- and 3D space respectively  and the
corresponding spectrum displays a characteristic distribution. As mentioned earlier, what is probably more
meaningful in respect of the boundary shape is the auto-correlation function of the level positions.
Practically, this requires that a given measurement picks out the full spectrum with extremely high
resolution. 
Unfortunately, this step is highly compromised by the inherent limitations of inverse
Laplace transform with which one transforms the time-domain data into the spectrum. In addition to
this, the preparation and detection states usually have high degree of symmetry, thus picking out only
a subset of the full spectrum. 

\vskip 0.08in
\centerline{{\bf APPLICATION}}
\vskip 0.08in
\noindent

\begin{figure}[p]
\includegraphics[width = 5 in]{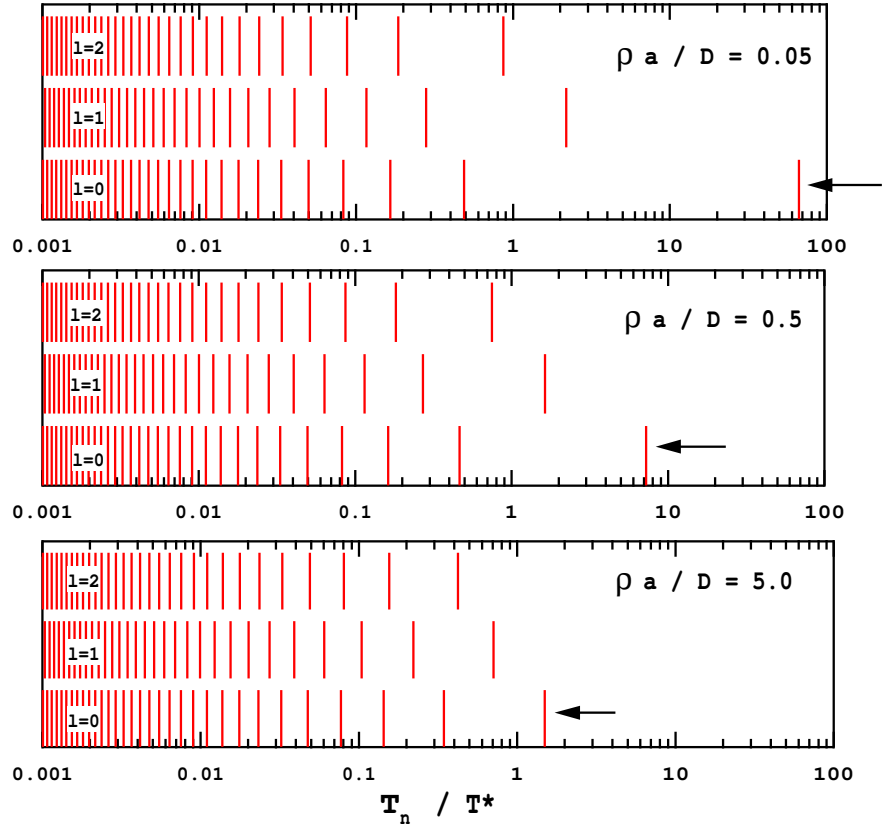} \\
\caption{Spectral variation for a sphere of radius $a$. The control parameter $\rho a / D$ is varied
from $0.05$ to $5.0$. For each value of this, three groups are shown with angular momentum $\ell = 0,
1, 2.$ The arrow marks the lowest eigen mode which shows an extreme sensitivity to $\rho.$
 }
\label{fig:sphere}
\end{figure}

Usually, the final state is detected by measuring the net magnetization. This is equivalent to
projecting the final state onto a uniform state.  If one also prepared a uniform initial state,
$|\Psi_i >  = | \Psi_0>,$ then we obtain 
\begin{equation}
{\cal M} (t) = < \Psi_0 | \Psi >_t =  \sum_n e^{-{E_n\over \hbar} t}  | <n | \Psi_0>|^2.
\end{equation}
For small $\rho$, we know from Equation \ref{eq:en2} that the lowest eigenmode becomes close to the
{\rm uniform} mode $\Psi_0 \sim {1 \over \sqrt{V}}$, with $E_0 \sim \rho S / V$ following directly
from the equation.  Furthermore, from orthogonality condition, we expect $<n|0> = \delta_{n0}$ and
therefore, 
${\cal M} (t) \sim e^{-{E_0\over \hbar} t}.
$ 
This immediately shows the limitations of detecting uniform magnetization as a probe for multiple length scales in a given pore. 
Even for determination of the average surface-to-volume ratio, ${\cal S} \over V$ from the observed 
$E_0 \sim \rho {{\cal S} \over V}$ is problematic unless the value of $\rho$ is independently and accurately
known since the {\em often dominant}, lowest eigenmode is extremely sensitive to $\rho$ and the 
all higher modes which remain robust as $\rho$ is varied, unfortunately do not show up in measurements. 
To illustrate this quantitatively, we show in Figure
\ref{fig:sphere} partial spectrum for a spherical pore of radius $a$ as we vary the control parameter 
$\rho a / D$ from $0.05$ to $5$.  
The general eigenmodes for a sphere is given by 
$\Psi_{k,l,m} = j_l (k r) Y_l^m$ with $l=0,1,\ldots$ and $m = -l, \ldots 0, \ldots l.$  For given $l$ and 
$m$, there is a manifold of eigenmodes with $k$ determied from 
\begin{equation}
k a {l j_{l-1} (k a) - (l+1) j_{l+1} (k a) \over 2 l + 1} +{\rho a \over D} j_l (ka ) = 0.
\end{equation}
Usually, only the subset with $l = m = 0$ is considered on symmetrical grounds with isotropic {\em initial} and
{\em detected} states $|\Psi_0>$, but for more
general cases (See discussion below), full angular variation should be retained even with the isotropic boundary
condition.  The most salient feature in the figure is the extreme sensitivity of the lowest eigenmode on $\rho$.
The higher modes, including the lowest modes in each $l\ne 0$ manifolds, are less sensitive and their positions
give direct information on the size of the sphere $a$. 
Another important characteristics of the higher modes is that they generally lead to a universal density of
states, {\em a la free electron model}, which reflects the fact that the most pore walls look very similar on
microscopic length scales, and the  fast decay modes are insensitive to the large scale variation of the pore
wall.

\begin{figure}[p]
\includegraphics[width = 7 in]{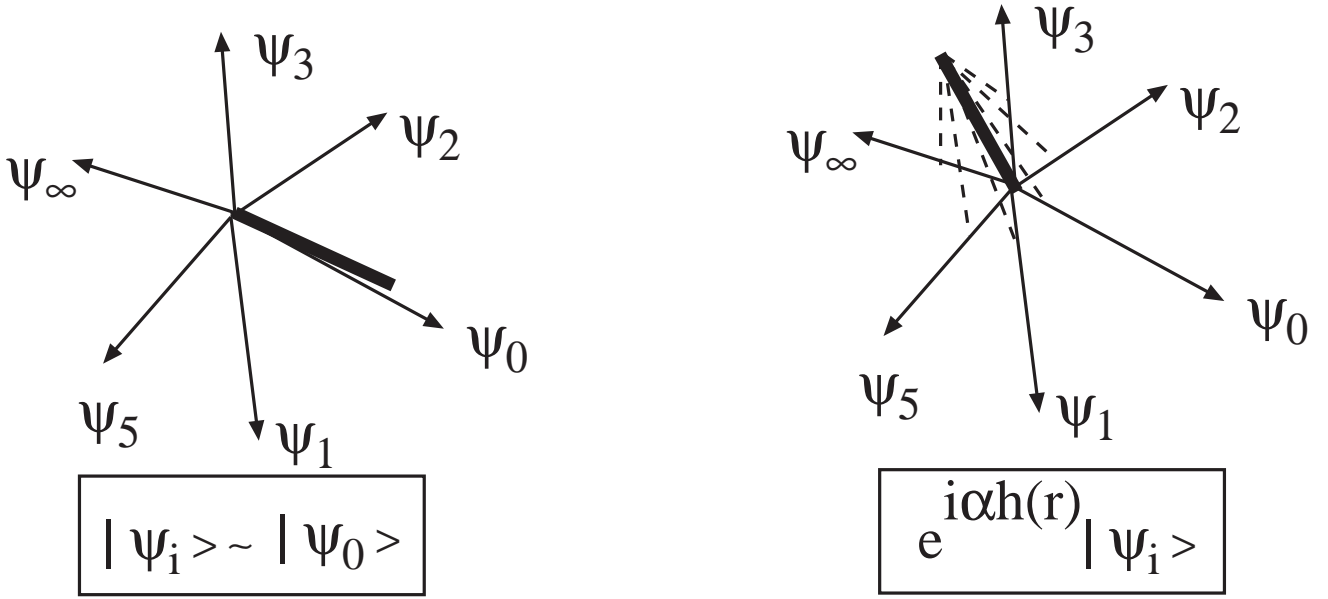} \\
\caption{Comparison between the probe using uniform initial \& final states (left) and 
that using states {\em rotated} in the Hilbert sub-space spanned by the eigenmodes conforming to the 
symmetry selection rule of the probe operator.
 }
\label{fig:hilbert}
\end{figure}

There are ways to get around the barrier to the excited states. One way is to create and detect a {\em
localized state}. This extreme case begins with a localized polarization 
$< {\bf r}| \Psi_i > = \delta ({\bf r} -
{\bf r}_0 )$, and  detecting again at the same spot, $< \Psi_i |$. The observed signal will follow: 
\begin{equation}
{\cal M}_{loc} ({\bf r}_0) = \sum_n e^{- t / T_n} |< n | {\bf r}_0 > |^2
\end{equation}
with the spectral weight $|< {\bf r}_0 | n>|^2$ for each mode democratically distributed over a broad range.
The two cases considered so far are two extreme limits where the {\em prepared} and the {\em detected} states
are oriented in the Hilbert space with some overlap with the (uniform) ground state of
the given boundary value problem, as schematically depicted in Figure \ref{fig:hilbert}. 
The most general situation may be described in terms of a unitary rotation of the uniform state by the
application of a local phase encoding-decoding operation, 
$e^{i\alpha ({\bf r})}$: 
\begin{equation}
{\cal M}_{\alpha}  = \sum_n e^{- t / T_n} |< n |e^{i \alpha ({\bf r})} | \Psi_0 > |^2.
\end{equation}
This operation can indeed be realized in the laboratory via application of $90$- and $180$-degree rf pulses,
a common MR techque. The critical ingredient for this to work is the presence of {\em internal field
inhomogeneity} on sub-pore  length scales. In the presence of a uniform applied field along the z-axis, the
small magnetic susceptibility contrast between the matrix and its filling fluid induces a local field
proportional to
$\triangle \chi.$ Therefore, one may rotate the spins onto the plane transverse to the polarizing field, and let
them precess  under the local field $h({\bf r})$ for a given time $\tau$, accumulating a phase $\phi({\bf r}) =
\gamma h ({\bf r})
\tau$. 

The spins are then rotated back, orthogonal to the transverse plane and allowed to relax over
time $t$ before one apply the reversing process. This winding-unwinding procedure would work perfectly if the 
molecules do not diffuse. However, the wound  (or encoded) state now evolves with 
$|n> e^{-t/ T_n} <n|$ with significantly enhanced overlap with all excited states $|n>$ with $n > 0.$
One can immediately see a potential benefit if the modulating field $h({\bf r})$ closely follows the underlying
geometry: The rotation of $|\Psi_i>$ will occur in such a way that the maximal overalp with the class
of eigenmodes with length scales matching the variation of the boundary. 
One can qualitatively understand this by noting that the local field can be described as arising from an
inhomogeneous distribution of surface dipoles whose strength is proportional to ${\bf h} \cdot {\hat n}$. 
Therefore, the local field tends to get enhanced in the vicinity of the pore wall, and is sensitive to its local
orientation. The {\em encoded} pattern created by flashing of the local field (in practice, the spins are
momentarily rotated to {\em feel} its presence) therefore reflects the contour of the pore wall in a convoluted
manner. Notwithstanding some detailed issues, this is quite sufficient in bringing out the full eigenspectrum
into light. 

Finally, we note that the widely studied $D(t) \propto{ |{\bf r}(t) - {\bf r}(0)|^2}$ is 
analogous to  the polarizability $\sum_n  <\Psi_f | \hat{\bf r} | n > e^{-t {\cal H}} < n | \hat{\bf
r} | \Psi_i > $ which for a rectangle leads to 
${1 \over 12} (L_x^2 + L_y^2 + L_z^2) + {1 \over \pi^2} \sum_{n>0} \sum_i
 e^{-t {D_0 \pi^2 4 n^2 \over L_i^2}} {L_i^2 \over n^2 } (-1)^n.
$ This, along with generic properties of the excited states, leads to an analytic for $D(t)$ which covers a wide
range of $t$ turning it into an effeftive probe into the pore shape as will be reported elsewhere.

\begin{figure}[p]
\includegraphics[width = 5 in]{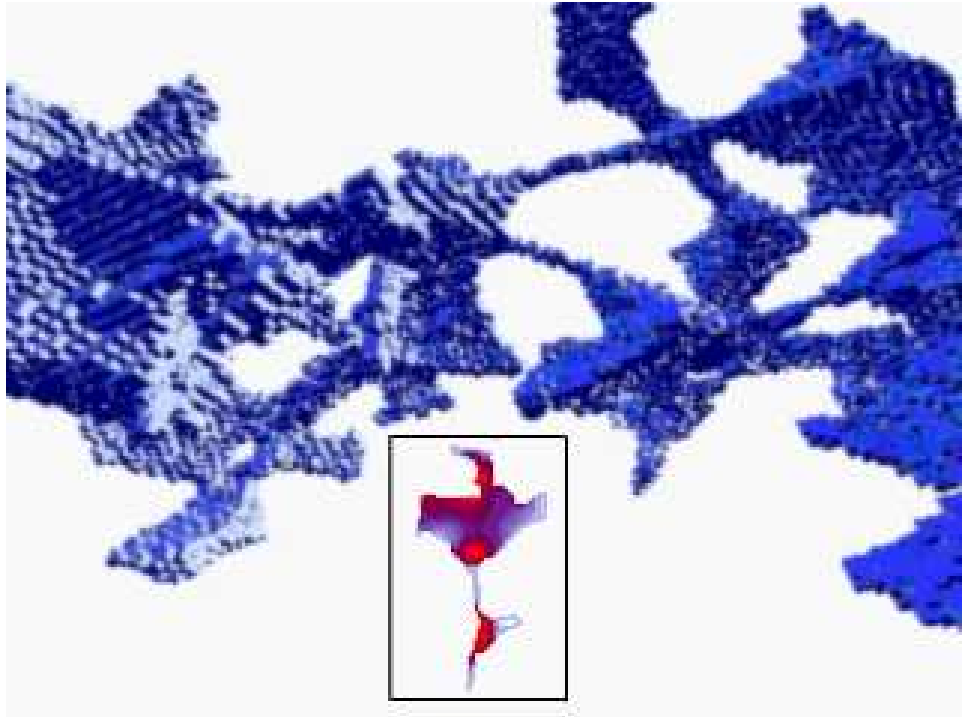} \\
\caption{A typical 3D pore strucure used in numerical simulation. The pore (light gray) is composed of about 
70000 connected cubes of size $5.72^3 \mu m^3$. The inset shows a typical cross section with color coding
representing the local field variation.
 }
\label{fig:FB}
\end{figure}

To be concrete, we are carrying out large scale numerical simulations of spin diffusion in various pore
structures such as random glass bead pack and real rocks obtained with 3D microtomographic technique.$^8$ 
The complex 3D pores are represented as coupled cluster of micron length scale cubes in which a large number of
random walkers (typically $10^4 -10^5$ in numbers) are deployed to simulate a variety of MR
measurements.  For ${\cal M}_{\alpha},$ we obtain the internal field by numerically evaluating dipole
contributions from all surrounding chambers. Looking at the results, an example of which is shown in Figure
\ref{fig:FB},  one can indeed verify the claim made earlier on the connection between the pore shape and the
local field configuration.

\begin{figure}[p]
\includegraphics[width = 5 in]{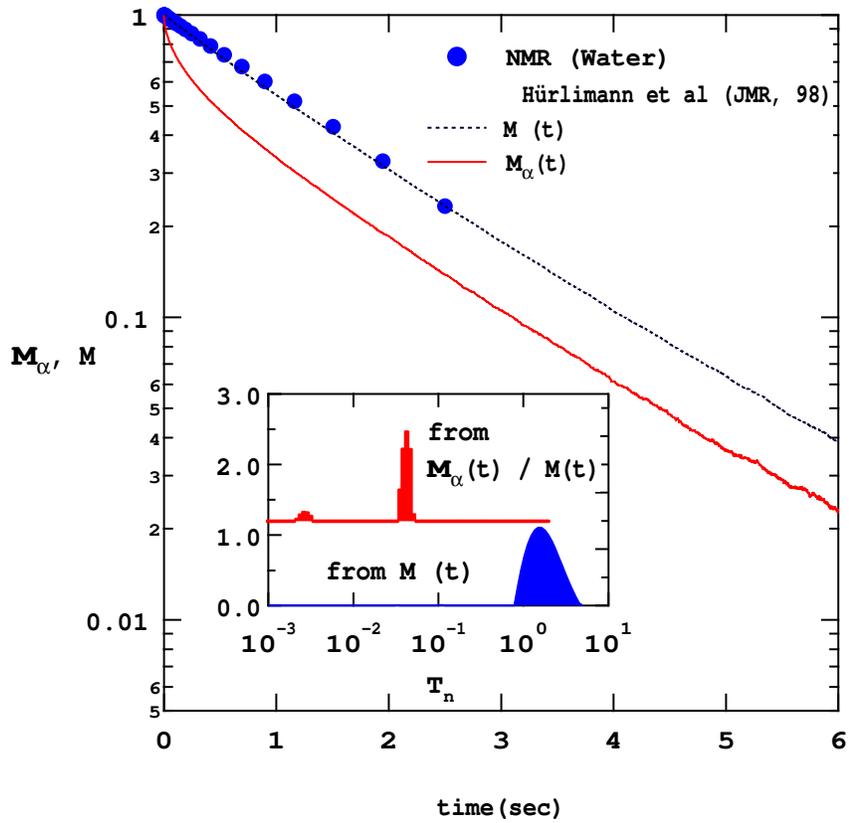} \\
\caption{Simulated ${\cal M}(t)$ and ${\cal M}_\alpha$ performed on the Fontainebleu sandstone structure. The
dots are from NMR experiment by H\"urlimann {\em et al}.$^7$ The inset shows the spectral weight derived
by performing an inverse Laplace transform. 
 }
\label{fig:ddif}
\end{figure}

Figure \ref{fig:ddif} shows a typical simulation results performed on the 3D microtomographic data of
Fontainebleu sample. The result for ${\cal M}$ shows excellent agreement with experimental data on the same
class of rocks. Furthermore, the simulated ${\cal M}_\alpha$ excavates the length scales hidden in ${\cal M}$ 
as shown in the inset for $|< n |e^{i \alpha ({\bf r})} | \Psi_0 > |^2$. The presence of multiple length
scales manifests itself in apprearence of the peaks, and their spectral weight depends on the strength of
$\alpha \propto \triangle \chi B \tau$. More detailed numerical studies on the variety of pore structure is under
progress and will be reported elsewhere.

\break

\break

\begin{figure}[p]
\includegraphics[width = 5 in]{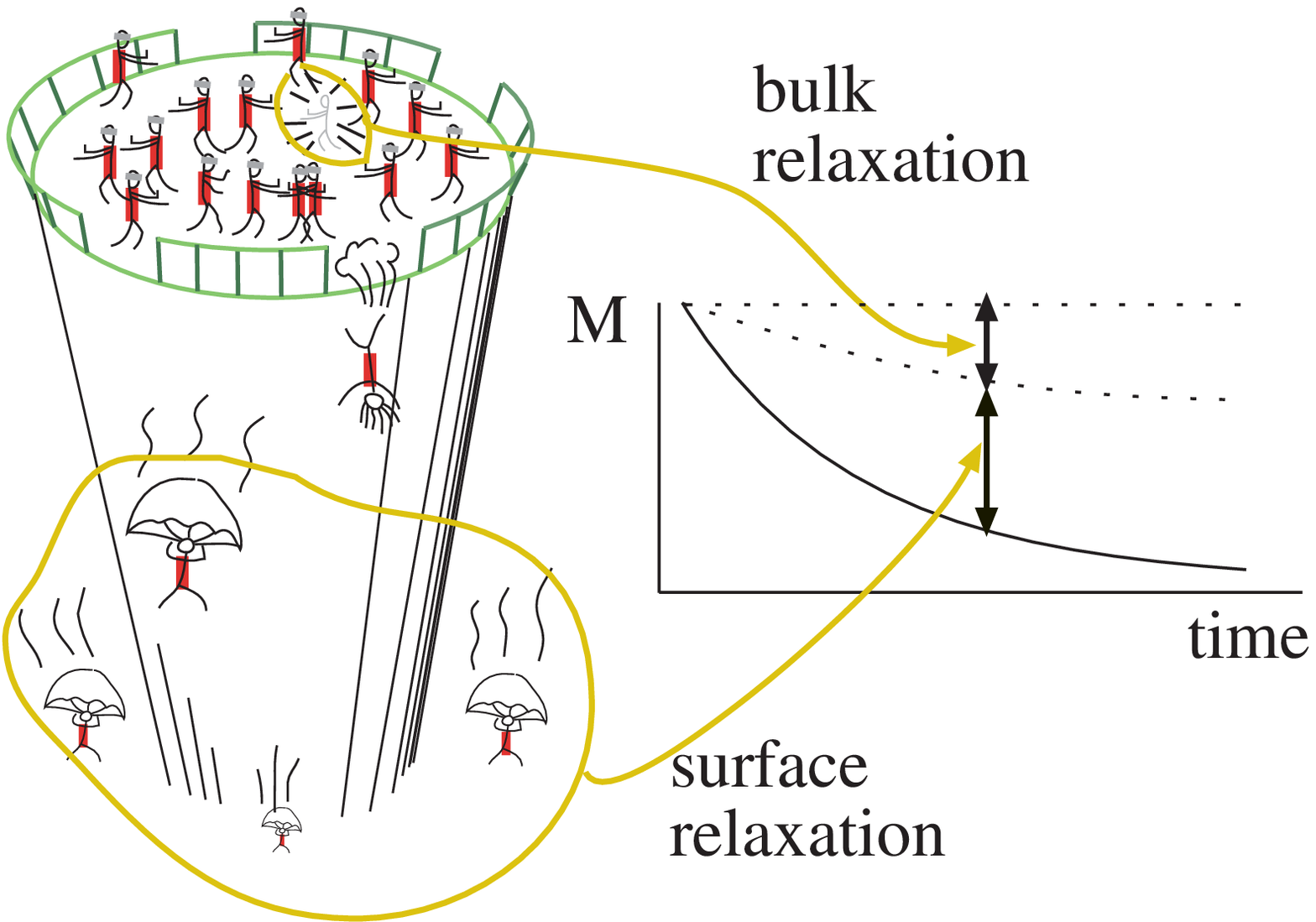} \\
\caption{Evolution of a population of random walkers subject to sinks on the edge. 
 }
\label{fig:zombie}
\end{figure}

\newpage

\newpage

\newpage

\end{document}